\documentclass[conference,letterpaper]{IEEEtran}

\usepackage[utf8]{inputenc} 
\usepackage[T1]{fontenc}
\usepackage{url}
\usepackage{ifthen}
\usepackage{cite}
\usepackage[cmex10]{amsmath} 

\usepackage{amssymb,amsfonts}
\usepackage{graphicx}

\newcommand{\bbC}{\mathbb{C}}\newcommand{\rmc}{\mathrm{c}}

\newcommand{\bbE}{\mathbb{E}}

\newcommand{\bbH}{\mathbb{H}}

\newcommand{\bbR}{\mathbb{R}}
\newcommand{\rms}{\mathrm{s}}

\newcommand{\bfA}{\mathbf{A}}\newcommand{\bfa}{\mathbf{a}}

\newcommand{\sfC}{\mathsf{C}}

\newcommand{\bfG}{\mathbf{G}}
\newcommand{\bfH}{\mathbf{H}}\newcommand{\sfH}{\mathsf{H}}
\newcommand{\bfI}{\mathbf{I}}
\newcommand{\bfJ}{\mathbf{J}}
\newcommand{\bfK}{\mathbf{K}}

\newcommand{\bfR}{\mathbf{R}}

\newcommand{\sfT}{\mathsf{T}}
\newcommand{\bfU}{\mathbf{U}}\newcommand{\bfu}{\mathbf{u}}
\newcommand{\bfV}{\mathbf{V}}

\newcommand{\bfX}{\mathbf{X}}\newcommand{\bfx}{\mathbf{x}}\newcommand{\sfX}{\mathsf{X}}
\newcommand{\bfY}{\mathbf{Y}}\newcommand{\sfY}{\mathsf{Y}}
\newcommand{\bfZ}{\mathbf{Z}}\newcommand{\sfZ}{\mathsf{Z}}

\newcommand{\I}{I}

\newcommand{\micnd}[3]{{\I}\left(\left. #1 ; #2 \,\right| #3\right)}

\newcommand{\expect}[1]{{\mathbb{E}}\left[#1\right]}

\newcommand{\variance}[1]{{\mathsf{Var}}\left[#1\right]}

\newcommand{\vect}[1]{{\text{vec}}\left(#1\right)}

\newcommand{\tr}[1]
{{\text{tr}}\left(#1\right)}

\begin{document}
\title{A Poincaré Lower Bound Approach for Performance Trade-offs in MIMO ISAC Systems with Blockage} 


\author{%
\IEEEauthorblockN{Mohammadreza Bakhshizadeh Mohajer$^{*}$, Luca Barletta$^{*}$,  Daniela Tuninetti$^{\dagger}$, Alessandro Tomasoni$^{**}$, \\ Daniele Lo Iacono$^{**}$, and Fabio Osnato$^{**}$} 
$^{*}$ Politecnico di Milano, 20133 Milano, Italy. Email: $\{$mohammadreza.bakhshizadeh, luca.barletta$\}$@polimi.it \\
$^{\dagger}$ University of Illinois Chicago, 60607 Chicago, IL, USA. Email: danielat@uic.edu \\
$^{**}$ STMicroelectronics Srl, 20864 Agrate B.za, Italy. Email: \texttt{firstname}.\texttt{lastname}@st.com}

\maketitle


\begin{abstract}
    Characterizing the performance trade-offs between sensing and communication subsystems is essential for enabling integrated sensing and communication systems. Various metrics exist for each subsystem; however, this study focuses on the ergodic capacity of the communication subsystem. Due to the complexity of deriving the sensing mean square error (MSE) and the inapplicability of the Bayesian Cramér-Rao Bound to channels with discrete or mixed distributions, this work proposes a Poincaré lower bound on the sensing MSE to address these issues.
    An achievable inner bound for the rate-sensing trade-off in a fading multiple-input multiple-output channel with additive white Gaussian noise and blockage probability is established. In addition, a strategy that is asymptotically optimal for sensing is provided.
\end{abstract}

\section{Introduction}
The 6G mobile communication system has emerged as a significant focus of global research, encompassing a range of innovative application scenarios, including smart cities, intelligent transportation, smart manufacturing, and smart homes~\cite{6G_EI}. Precise sensing is expected to evolve from an auxiliary function to a fundamental service among the notable features of 6G, thereby adding an additional dimension of capability to the network~\cite{towards_6g,ISAC_survey,5G_6G_ISAC}. 
There are various performance metrics for each individual system and based on their perspective. On one hand, the communication system is designed to maximize the information rate using an optimal transmitted signal while having knowledge of the channel. On the other hand, the sensing subsystem solves the problem of estimating the parameters of the channel while it has knowledge of the transmitted signal~\cite{ISAC_Caire}. Future 6G vehicle-to-everything (V2X) is foreseen to explore millimeter wave (mmWave)~\cite{V2X}. Furthermore, the anticipated utilization of millimeter-wave (mmWave) and terahertz (THz) frequencies for sensing in 6G and next-generation wireless networks is foreseen~\cite{5G_6G_ISAC}. These frequencies are subject to increased losses and are more susceptible to blockage~\cite{UAV_blockage}.

\subsection{Previous Work}
Motivated by integrated sensing and communication (ISAC) systems and propagation challenges in high frequencies, we study the performance trade-off between the two subsystems. The performance of the communication subsystem is typically evaluated by the ergodic capacity. Due to the challenging evaluation of the mean square error (MSE), which is the typical performance metric of the sensing subsystem, the Bayesian Cramér-Rao bound (BCRB) on the MSE has been adopted by previous work~\cite{ISAC_Caire}. 
In contrast to the classic Cramér-Rao bound (CRB), which depends on the actual value of the parameter being estimated, the BCRB utilizes the \textit{a priori} distribution of the parameter and provides a global bound~\cite{van_trees_BCRB}. However, the issue with BCRB is that it is generally not tight and does not apply to scenarios involving discrete or mixed distributions \cite{Poincare_dytso}. To address this issue, in our previous work~\cite{WCNC_conf}, we proposed an alternative lower bound on the MSE of channel estimation based on a Poincaré-type inequality derived in~\cite{Poincare_dytso}. Specifically, in~\cite{WCNC_conf} we computed a lower bound on the sensing MSE for multiple-input multiple-output (MIMO) additive Gaussian channels with Rayleigh fading and blockage events. 

\subsection{Contributions}
In this work, we compute an achievable sensing-rate region along with an outer bound for a MIMO ISAC system. Furthermore, we suggest an achievable strategy that is asymptotically sensing optimal for high sensing signal-to-noise ratio (SNR). Finally, through numerical simulations, we evaluate the inner bound and discuss the results.

\subsection{Paper Organization}
In Sec.~\ref{Section:model_innerbound}, we introduce the system model and the performance metrics for both sensing and communication subsystems and we define the structure of sensing-rate inner bound. The achievable strategies are studied in Sec.~\ref{Section:Optimal_region}. The numerical evaluation of the achievable boundary and a discussion on the results are provided in Sec.~\ref{Section:result_discussion}. Finally, Sec.~\ref{Section:conclusion} concludes the paper.

\subsection{Notation}
Throughout the paper, deterministic scalar quantities are denoted by lowercase letters, random vectors are denoted by uppercase sans serif letters, deterministic vectors are denoted by bold lowercase letters, and random matrices by bold uppercase letters (e.g. $x, \sfX, \bfx, \bfX$). $\langle \cdot, \cdot\rangle$ denotes the inner product. For a matrix $\bfA$, $\bfA^{\sfT}$, $\bfA^{\dagger}$, $\bfA^{-1}$, $|\bfA|$, and $\text{tr}(\bfA)$ denote the transpose, the Hermitian transpose, the inverse, the determinant and the trace of the matrix $\bfA$, respectively. $(\bfA)^{+} = (\bfA^{\sfT} \bfA)^{-1} \bfA^{\sfT}$ denotes the pseudoinverse of matrix $\bfA$. For matrix $\bfA \in \bbR^{n\times m}$, $\text{vec}(\bfA) = [\bfa_{1}^{\sfT}, \bfa_{2}^{\sfT}, \dots , \bfa_{m}^{\sfT}]^{\sfT} \in \bbR^{nm \times 1}$, where $\bfa_i$ is the $i$-th column of $\bfA$, is the vectorization operator. $\bfI_{k}$ is the identity matrix of dimension $k$, $\bf0$ is the column vector of all zeros, $\Gamma(\cdot)$ indicates the gamma function, and $\delta(\cdot)$ denotes the Kronecker delta. The smallest eigenvalue and the smallest singular value are denoted by $\lambda_{\textrm{min}}(\bfA)$ and $\sigma_{\min}(\bfA)$, respectively. $\succeq$ denotes the semidefinite ordering and the Kronecker product is denoted by $\otimes$. $\nabla_{\bfx}(\cdot)$, $\bfJ_{\bfx}(\cdot)$, and $\bbH_{\bfx}(\cdot)$ denote the gradient, the Jacobian matrix, and the Hessian matrix with respect to $\bfx$. The minimum MSE (MMSE) of estimating the random vector $\sfX$ given observation $\sfY$ is defined as
\begin{equation}\label{eq:definition_MMSE}
    \text{mmse}(\sfX|\sfY) \triangleq \bbE \left[\|\sfX-\bbE[\sfX|\sfY]\|^{2} \right],
\end{equation}
where $\bbE[\cdot]$ denotes the expectation operator and $\|\cdot \|$ is the Euclidean norm. The $\variance{\cdot}$ is defined as
\begin{align}
    \variance{\sfX} &\triangleq \bbE \left[\|\sfX-\bbE[\sfX]\|^{2} \right].
\end{align}
We also consider the following complex to real mapping for column vectors and matrices 
\begin{align} \label{eq:ReIm_matrix}
\overline{\bfx} = 
    \begin{bmatrix}
        \text{Re}({\bfx}) \\
        \text{Im}({\bfx})
    \end{bmatrix},\,
    \overline{\bfX} = 
    \begin{bmatrix}
        \text{Re}({\bfX}) & -\text{Im}({\bfX}) \\
        \text{Im}({\bfX}) & \text{Re}({\bfX})
    \end{bmatrix}.
\end{align}

\section{System Model and Achievable Inner Bound}\label{Section:model_innerbound}
\subsection{System Model}
Let us consider the general MIMO ISAC system model
\begin{align}
    \bfY_{i} &= \bfH_{i} \bfX + \bfZ_{i}, \quad i \in \{ \rm c ,\rm s \} \label{eq:sensing_sys_model}
\end{align}
where $\bfY_{\rm c} \in \bbC^{N_{\rm c} \times T}$ and $\bfY_{\rm s} \in \bbC^{N_{\rm s} \times T}$ are the communication and sensing received signals, $\bfX \in \bbC^{M \times T}$ is the transmitted dual-function signal for both communication and sensing tasks, $\bfH_{\rm c} \in \bbC^{N_{\rm c} \times M}$ denotes the communication channel matrix, and $\bfH_{\rm s} \in \bbC^{N_{\rm s} \times M}$ is the target response matrix. The channel matrices have the following fading distribution 
\begin{align}
f_{\bfH_{i}}(\bfx) = (1-\alpha_{i}) \delta(\bfx) + \alpha_{i} \mathcal{CN}\left(\bfx;{\bf0},\sigma_{\sfH_{i}}^2 \bfI_{N_{i}M}\right), i \in\{ \rmc,\rms \}
\end{align}
where $1-\alpha_{i}$, for $\alpha_{i} \in (0,1]$, is the probability that receiver~$i$ experiences a blockage event that lasts $T$ channel uses;
both channels vary every $T$ symbol in an independent and identically distributed (i.i.d.) manner; and the noise vectors~$\bfZ_{i}$ are assumed to have i.i.d. circularly symmetric, zero mean, complex Gaussian components with variance $\sigma_{i}^2$, for $i \in\{ \rmc,\rms \}$.

The sample covariance matrix $\bfR_{x}$ of the transmitted signal will play a central role in both the analysis of the sensing system's performance and in imposing an average-power constraint $P_{0}$ for the communication problem, i.e., 
\begin{align} 
\tr{ \expect{\bfR_{x}} } \leq M P_{0}, \quad\bfR_{x} \triangleq T^{-1}\bfX \bfX^{\dagger}\in \bbC^{M \times M}.
\label{eq:APowerConstraint}
\end{align} 

We assume $\bfX$ to be known at the sensing receiver, which observes $\bfY_{\rm s}$ and aims to estimate the target response matrix $\bfH_{\rm s}$. The communication subsystem aims to transmit as much information as possible through optimal $\bfX$ while the transmitter and the communication receiver have full knowledge of $\bfH_{\rm c}$. The performance of the communication subsystem is characterized by the ergodic rate  
\begin{align} 
    R  &\triangleq T^{-1} \micnd{\bfY_{\rm c}}{\bfX}{\bfH_{\rm c}}.
    \label{eq:Achievable_Rate}
\end{align}

The sensing performance is the MSE
\begin{align}
    \epsilon &\triangleq \expect{\|\bfH_{\rm s} - \widehat{\bfH}_{\rm s} \|^2},
    \label{eq:MSE of A def}
\end{align}
where $\widehat{\bfH}_{\rm s}$ is an estimate of $\bfH_{\rm s}$ based on $(\bfY_{\rm s}, \bfX)$.

The sensing-rate region $\mathcal{C}$ consists of all feasible pairs $(\epsilon, R)$~\cite{ISAC_Caire}. The Pareto front~\cite{Pareto_boundary} of $\mathcal{C}$ can be determined as
\begin{align}
    \min_{p_{\bfX} \in \cal{F}}\,\, \epsilon - \lambda T^{-1}\micnd{\bfY_{\rm c}}{\bfX}{\bfH_{\rm c}},
    \label{eq:paretoFrontierOfC}
\end{align}
where $\lambda \in [0,\infty)$ controls the preference between the sensing and the communication performance, and $\cal{F}$ is the feasibility region of $p_{\bfX}$ determined by~\eqref{eq:APowerConstraint}.

\subsection{Time-Sharing Achievable Region}
An achievable sensing-rate region, following~\cite{ISAC_Caire}, is as follows:
\begin{align}
    \epsilon &\geq \epsilon_{\textrm{min}},\\
    R &\leq R_{\textrm{max}},\\
    \epsilon &\geq \epsilon_{\textrm{min}} + \frac{\epsilon_{\textrm{CommOpt}} - \epsilon_{\textrm{min}}}{R_{\textrm{max}} - R_{\textrm{SenseOpt}}} (R - R_{\textrm{SenseOpt}}),
\end{align}
where
\begin{align}
    \epsilon_{\textrm{min}} &:= \min_{p_{\bfX}\in \cal{F}} \epsilon,\\
    R_{\textrm{max}} &:= \max_{p_{\bfX}\in \cal{F}} T^{-1} \micnd{\bfY_{\rm c}}{\bfX}{\bfH_{\rm c}},\\
    \epsilon_{\textrm{CommOpt}} &:= \min_{p_{\bfX}\in \cal{F}} \epsilon,\quad \text{s.t.}\quad T^{-1} \micnd{\bfY_{\rm c}}{\bfX}{\bfH_{\rm c}} = R_{\textrm{max}},\\
    R_{\textrm{SenseOpt}} &:= \max_{p_{\bfX}\in \cal{F}} T^{-1} \micnd{\bfY_{\rm c}}{\bfX}{\bfH_{\rm c}},\quad \text{s.t.}\quad \epsilon = \epsilon_{\textrm{min}}.
\end{align}

Two important points of the sensing-rate region are the sensing-optimal and the communication-optimal points which are defined as 
$
(\epsilon_{\textrm{min}}, R_{\textrm{SenseOpt}})$, and
$
(\epsilon_{\textrm{CommOpt}}, R_{\textrm{max}})$,
respectively. The line connecting these two points can be achieved using the time-sharing strategy~\cite{Elements_cover}.

In the next section, we start by characterizing the two optimal points and propose waveforms that can achieve them. We then also propose achievable and converse bounds for~\eqref{eq:paretoFrontierOfC}.

\section{Sensing and Communication Optimal Points}\label{Section:Optimal_region}

By standard arguments, it can be shown that~\eqref{eq:sensing_sys_model} can be rewritten in vector form as
\begin{align}
    \overline{\vect{\bfY_{\rm s}^\sfT}} = (\bfI_{N_{\rm s}} \otimes \overline{\bfX^\sfT})  \overline{\vect{\bfH_{\rm s}^\sfT}} +  \overline{\vect{\bfZ_{\rm s}^\sfT}}. \label{eq:reim_vect_sensing}
\end{align}
Let us introduce the quantities $\sfY_{\rm s} \triangleq \overline{\vect{\bfY_{\rm s}^\sfT}}$, $\sfH_{\rm s} \triangleq \overline{\vect{\bfH_{\rm s}^\sfT}}$, $\sfZ_{\rm s} \triangleq \overline{\vect{\bfZ_{\rm s}^\sfT}}$, and $\sfC_{\bfX} \triangleq \bfI_{N_{\rm s}} \otimes \overline{\bfX^\sfT}$ so that the complex-valued vectorized sensing model \eqref{eq:reim_vect_sensing} becomes
\begin{align} \label{eq:new_notation_model}
    \sfY_{\rm s} = \sfC_{\bfX} \sfH_{\rm s} + \sfZ_{\rm s}.
\end{align}

In our previous work~\cite{WCNC_conf}, we derived a Poincaré-type lower bound on the MMSE of $\sfH_{\rms}$. 
Specifically, we demonstrated that in the high SNR regime the proposed bound on the sensing MMSE is a simple function of the sample covariance matrix~$\bfR_{x}$:
\begin{align}
    &\lim_{\sigma_{\rm s} \to 0} \frac{\text{mmse}(\sfH_{\rm s}|\sfY_{\rm s})}{\sigma^2_s}
    \geq \frac{\alpha_{\rm s}}{2} \bbE\left[\left( \text{tr} \left({\bfR_{x}}\right)\right)^{-1}\right].
    \label{eq:asympotot_slop}
\end{align}

\subsection{Sensing-Optimal Achieving Strategy}
In~\cite{ISAC_Caire}, it was shown  that for an AWGN channel with known transmitted signal $\bfX$ at the sensing receiver, the optimal sensing performance is achieved when the trace of the sample covariance matrix is deterministic, i.e., $\tr{\bfR_{x}} = \tr{\expect{\bfR_{x}}}$. Considering the high sensing SNR regime, the sensing-optimal ${\bfR_{x}}$ is the one that minimizes the right-hand side (RHS) of~\eqref{eq:asympotot_slop}. So,
\begin{align}\label{eq:optimal_Rx_vect}
    &\min_{p({\bfR_{x}})} \alpha_{\rm s} \frac{\sigma_s^2}{2} \bbE\left[\left( \text{tr} \left({\bfR_{x}}\right)\right)^{-1}\right],\\
    &\text{s.t.} \quad \text{tr}\left(\expect{{\bfR_{x}}}\right) = P_{0} M, {\bfR_{x}} \succ 0, {\bfR_{x}}=\bfR_{x}^{\dagger}.
\end{align}
We observe that $\frac{1}{\text{tr} \left({\bfR_{x}}\right)}$ is convex, then by applying Jensen's inequality we have
\begin{align}\label{eq:Jensen_sensing}
    \min_{\bfR_{x}} \alpha_{\rm s} \frac{\sigma_s^2}{2} \bbE\left[(\text{tr}\left({\bfR_{x}}\right))^{-1}\right] \geq \min_{\bfR_{x}} \alpha_{\rm s} \frac{\sigma_s^2}{2} \bbE\left[\text{tr}\left({\bfR_{x}}\right)\right]^{-1}.
\end{align}
The RHS of~\eqref{eq:Jensen_sensing} is minimized when $\bbE\left[\text{tr}\left({\bfR_{x}}\right)\right] = P_{0}M$ and the inequality holds with equality when $\text{tr}\left({\bfR_{x}}\right)$ is deterministic. As a result, the optimal solution of~\eqref{eq:optimal_Rx_vect} is when $\tr{\bfR_{x}} = P_{0}M$.

The corresponding communication rate $R_{\textrm{SenseOpt}}$ can be expressed as
\begin{align}
    R_{\textrm{SenseOpt}} = \max_{p_{\bfX}} T^{-1} \micnd{\bfY_{\rm c}}{\bfX}{\bfH_{\rm c}}, \ \text{s.t.} \ \tr{\bfR_{x}} = P_{0}M.
\end{align}
Consider the singular value decomposition of $\bfX$
\begin{align}
    \bfX = \sqrt{T} \bfU \bf{\Sigma} \bfV^{\dagger},
    \label{eq:StructureOfXopt}
\end{align}
where $\bfU \in \bbC^{M\times M}$ is the matrix of left singular vectors, $\bf{\Sigma}$ is the $M\times T$ matrix of singular values, and $\bfV \in \bbC^{T\times T}$ is the matrix of right singular vectors. As demonstrated in~\cite{Capacity_bound}, mutual information does not depend on $\bfV$. Next we propose a scheme for $\bfX$ by setting the diagonal elements of $\bf{\Sigma}$ equal to $\sigma_{1} = \sigma_{2} = \dots = \sigma_{M} = \sqrt{ M P_{0}}$ and we select $\bfU$ as an isotropically distributed (i.d.) unitary matrix. The probability density function of an i.d. matrix is invariant under left-multiplication by any deterministic unitary matrix~\cite{USTM}. According to~\cite{Capacity_bound}, an i.d. unitary matrix can be obtained by first generating a $T \times T$ random matrix $\bfA$ whose elements are independent ${\cal CN}(0,1)$, then performing the QR factorization of $\bfA = \bfU \bfG$, where $\bfU$ is unitary and $\bfG$ is upper triangular, resulting in $\bfU = \bfA \bfG^{-1}$. Note that $\bfX$ has deterministic trace of the sample covariance by construction. 

Following the proposed scheme, we obtain a lower bound on $R_{\textrm{SenseOpt}}$ as follows~\cite[Eq.~(10)]{Capacity_bound}:
\begin{align}
    R_{\textrm{SenseOpt}}&\ge \alpha_{\rm c} T^{-1}\left[ -TN_{\rm c} \log e - N_{\rm c} M \log{\left(1 + \frac{P_{0}T}{\sigma_{\rm c}^{2}}\right)} \nonumber\right. \\
    &\quad - \int d\lambda P(\lambda) f_{l}(\lambda) \nonumber\\
    & \quad \left.\left[ \log f_{l}(\lambda) - (\log e) \sum_{l=1}^{\text{min}(N_{\rm c},T)} \lambda_{l} \right] \right],
\end{align}
where $\lambda_{l}, l\in \{1,2,\dots,\text{min}(N_{\rm c},T)\}$ are the eigenvalues of $\bfY_{\rm c}\bfY_{\rm c}^{\dagger}$, $p(\lambda)$ is
\begin{align}
    &p(\lambda) =\\
    &\frac{e^{-\left( \sum_{l=1}^{\text{min}(N_{\rm c},T)}\lambda_{l} \right)} \left( \Pi_{l=1}^{\text{min}(N_{\rm c},T)} \lambda_{l} \right)^{|T-N_{\rm c}|} \Pi_{i<j}(\lambda_{i}-\lambda_{j})^{2}}{\Pi_{l=1}^{\text{min}(N_{\rm c},T)} \Gamma(T-l+1) \Gamma(N_{\rm c} -l+1) },
\end{align}
and $f_{l}(\lambda)$ is
\begin{align}
    f_{l} (\lambda) &= \left( 1+ \frac{P_{0}T}{\sigma_{\rm c}^{2}} \right)^{-N_{\rm c}M} \int d\bfu p_{\bfU}(\bfu) \nonumber\\
    & \exp{\left\{ \sum_{n=1}^{\text{min}(N_{\rm c},T)}\sum
    _{m=1}^{M} \lambda_{n} \left(\frac{P_{0} T}{\sigma_{\rm c}^{2} + P_{0}T} \right) |u_{nm}|^{2}\right\}}.
\end{align}
When $T \rightarrow \infty$, we can use the result for the information rate~\cite[Sec.V-C]{Capacity_bound}
\begin{align}
    R_{\textrm{SenseOpt}} = \expect{\alpha_{\rm c} \log \left| \bfI_{{N}_{\rm c}} + \frac{P_{0}}{\sigma_{\rm c}^{2}} \bfH_{\rm c}^{\dagger} \bfH_{\rm c} \right|}.
\end{align}

\subsection{Communication-Optimal Achieving Strategy}
The capacity-achieving strategy in the communication-only scenario is 
\begin{align}
    R_\textrm{max} &= \max_{p_{\bfX}} \frac{\alpha_{\rm c}}{T} \micnd{\bfY_{\rm c}}{\bfX}{\bfH_{\rm c}}, \quad\text{s.t.}\quad  \tr{ \expect{\bfR_{x}} } \leq M P_{0} \\
    &= \expect{\alpha_{\rm c} \log \left|\bfI + \sigma_{\rm c}^{-2} \bfH_{\rm c} \bfK_{\bfX}^{\textrm{(wf})} \bfH_{\rm c}^{\dagger} \right|},
\end{align}
where $\bfK_{\bfX}^{\textrm{(wf)}} =  \expect{\bfR_{x}^{\textrm{CommOpt}}}$ is the water-filling optimal covariance matrix given by $\bfK_{\bfX}^{\textrm{(wf)}} = \bfV {\bf\Lambda} \bfV^{\dagger}$. Then, each column of $\bfX$ follows a circularly symmetric complex Gaussian distribution ${\cal CN}(\bf0, \bfK_{\bfX}^{\textrm{(wf)}})$. The diagonal elements of $\bf{\Lambda}$ are given by
\begin{align}
    {\bf\Lambda}_{ii} = \max\left\{  \mu - \frac{\sigma_{\rm c}^{2}}{\eta_{i}^{2}} , 0 \right\},
\end{align}
where $\eta_{i}, \, i \in \{1,2,\dots,r\}$, are the $r$ nonzero singular values of $\bfH_{\rm c}$, and $\mu$ is the water level that can be obtained based on $\sum_{i} {\bf\Lambda}_{ii} = P_{0}M$~\cite{Pareto_boundary}.
Here, a ``short-term power constraint" is imposed, which is inherently stricter than a ``long-term power constraint." This approach facilitates solving the problem for each channel realization $\bfH_{\rm c}$ within each coherence time interval, ensuring feasibility for achievability~\cite{power_constraint}. 
Finally, the communication-constrained Poincaré lower bound is given by
\begin{align}
    \epsilon_{\textrm{CommOpt}} = \alpha_{\rm s} \frac{\sigma_s^2}{2} \bbE\left[\left( \text{tr} \left({\bfR_{x}^{\textrm{CommOpt}}}\right)\right)^{-1}\right].
\end{align}

\subsection{Converse Bound}
Based on the sensing and communication optimal points, we have the trivial converse bounds $R \leq R_\textrm{max}, \ \epsilon \geq \epsilon_\textrm{min}$.

\section{Numerical Results and Discussion}\label{Section:result_discussion}

In the following numerical evaluations, we set the model parameters as in Table~\ref{tab:params}.
\begin{table}[htbp]
\caption{Parameters for target response matrix estimation.}
\label{tab:params}
\begin{center}
\begin{tabular}{|c|c|}
\hline
\textbf{Parameter}&{\textbf{Value}} \\
\hline
Number of Tx antennas ($M$) & $4$ \\
\hline
Number of sensing Rx antennas ($N_{\rm s}$) & $4$ \\
\hline
Number of communication Rx antennas ($N_{\rm c}$) & $4$ \\
\hline
Channel coherence period ($T$) & $16$ \\
\hline
Sensing blockage probability ($1-\alpha_{\rm s}$)& $0.6$ \\
\hline
Communication blockage probability ($1 - \alpha_{\rm c}$)& $0$ \\
\hline
Sensing transmit SNR ($P_{0}/\sigma_{\rm s}^{2}$) & $24$dB \\
\hline
Communication transmit SNR ($P_{0}/\sigma_{\rm c}^{2}$) & $24$dB \\
\hline
Sensing channel variance ($\sigma_{\sfH_{\rm s}}^{2}$) & $M^{-1}$ \\
\hline
Communication channel variance ($\sigma_{\sfH_{\rm c}}^{2}$) & $1$ \\
\hline
\end{tabular}
\label{tab:parameters}
\end{center}
\end{table}

\begin{figure}
    \centering
    \includegraphics[width=0.48\textwidth]{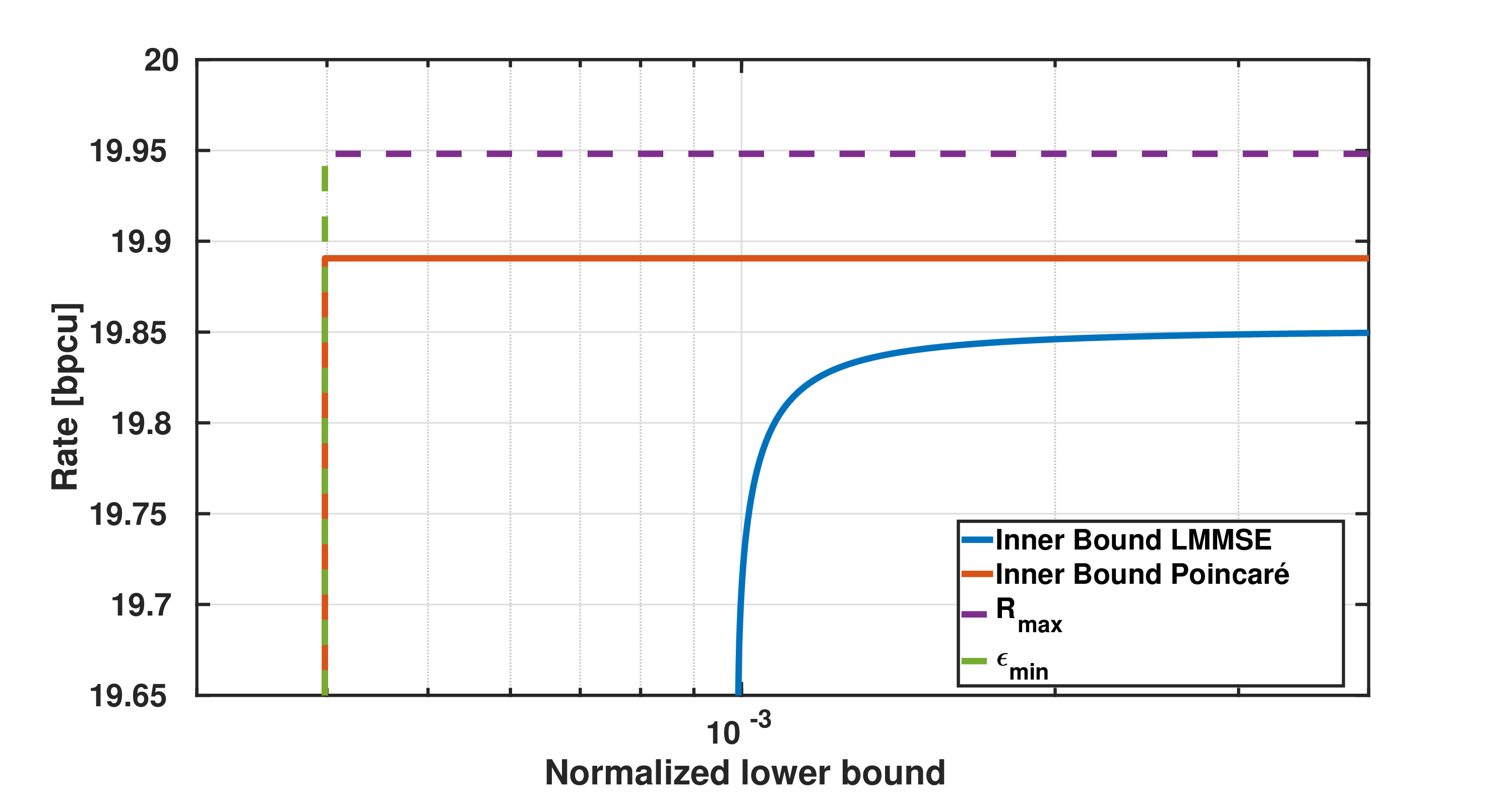}
    \caption{Inner bound and outer bound of the sensing-rate region for the task of target response matrix estimation, where both communication and sensing channel are subject to uncorrelated Rayleigh fading and blockage probabilities  $1-\alpha_{\rm c} = 0$ and $1-\alpha_{\rm s} = 0.6$.}
    \label{fig:pareto_bound}
\end{figure}
We compute an achievable region by solving 
\begin{align}
    &\max_{\bfR_{x}} \lambda \expect{\alpha_{\rm c} \log \left| \bfI + \sigma_{\rm c}^{-2} \bfH_{\rm c} \bfR_{x} \bfH_{\rm c}^{\dagger} \right|} \nonumber\\
    & \text{s.t.} \quad \tr{\expect{\bfR_{x}}} = P_{0}M, \, \bfR_{x} \succeq 0, \, \bfR_{x} = \bfR_{x}^{\dagger}, \nonumber\\
    & \alpha_{\rm s} \frac{\sigma_s^2}{2} \bbE\left[\left( \text{tr} \left({\bfR_{x}}\right)\right)^{-1}\right] \leq \epsilon_{\alpha},
\end{align}
for each channel realization $\bfH_{\rm c}$, where $\lambda \in [0,\infty)$ controls the power allocation strategy. Figure~\ref{fig:pareto_bound} shows the bound that is computed for each optimal value of $\bfR_{x}(\lambda)$ using the Poincaré lower bound as
\begin{align}
    R(\lambda) &= \expect{\alpha_{\rm c} \log \left|\bfI + \sigma_{\rm c}^{-2} \bfH_{\rm c} \bfR_{x}(\lambda) \bfH_{\rm c}^{\dagger} \right|},\label{eq:rate_bound} \\
    \epsilon(\lambda) &= \alpha_{\rm s} \frac{\sigma_{\rm s}^2}{2} \bbE\left[\left( \text{tr} \left({\bfR_{x}(\lambda)}\right)\right)^{-1}\right].
\end{align}
For comparison, we have also computed the inner bound based on the linear MMSE (LMMSE) for each optimal value of $\bfR_{x}(\lambda)$ as
\begin{align}\label{eq:LMMSE_vector_case}
    \epsilon_{\text{lmmse}}(\lambda) &= \frac{N_{\rm s} \sigma^2_s}{2T} \bbE\left[ \text{tr}\left(\left(\bfR_{x}(\lambda)+\frac{\sigma^2_s}{\alpha_{\rm s} T\sigma^2_{\sfH_{\rm s}}}\bfI\right)^{-1}\right) \right],
\end{align}
and the information rate using~\eqref{eq:rate_bound}.
Finally, we have normalized the value of the LMMSE as
\begin{align}
    \epsilon_{\text{normalized}} = \frac{\epsilon_{\text{lmmse}}(\lambda)}{MN_{\rm s}}.
\end{align}
It is worth noting that the LMMSE coincides with the MMSE for Gaussian random variables; however, for non-Gaussian pdfs, as in our case, it serves as an upper bound to the MMSE.

\subsection{Discussion}
From \figurename~\ref{fig:pareto_bound} we can see that the inner bound region is almost rectangular. This is due to the assumption $T \rightarrow \infty$. This assumption enabled us to greatly simplify the information rate expression, and therefore to find the information rate that satisfies the constraint on the trace of $\bfR_x$. Notice that, as $T \rightarrow \infty$, the distribution of the entries of $\bfX$ approach independent ${\cal CN}(0,1)$~\cite{USTM}, which is the  distribution that achieves the communication-optimal point. Therefore, both sensing-optimal and communication-optimal points have the same value and they collapse on each other resulting in a rectangular region. As a future research direction, it is interesting to address the case of finite $T$.

\section{Conclusion}\label{Section:conclusion}
In this work, we studied the performance trade-off between the sensing and communication subsystems of a multiple-input and multiple-output (MIMO) integrated sensing and communication (ISAC) system for an additive Gaussian channel with Rayleigh fading and blockage events, which is of great significance for next-generation wireless networks operating at millimeter wave. In such channels, the Bayesian Cramér-Rao bound (BCRB) is not applicable, because the a-priori distribution of the quantity to estimate is mixed discrete-continuous. Hence, we used a novel Poincaré-type lower bound on the sensing mean square error for the sensing part while the performance metric of the communication system is the ergodic capacity. Based on this, we proposed an achievable sensing-rate region along with an outer bound. We suggested an input distribution that is asymptotically optimal for sensing and studied its properties in the high signal-to-noise ratio regime and channel coherence time $T$ approaching infinity. We observed that for the case of large values of $T$, since the transmitted signal approaches an independent and identically distributed circularly symmetric Gaussian distribution, the two corner points of the achievable region collapse on each other. This motivates the expansion of future research to examine the information rate for finite \( T \) and to explore the optimal signaling strategy in this context. Another promising research path is to study the ISAC system for the case where the channel matrix is a nonlinear function of a hidden parameter, such as in angle of arrival estimation, when BCRB cannot be calculated.

\bibliographystyle{IEEEtran}
\bibliography{bibliography.bib}

\begin{thebibliography}{10}
\providecommand{\url}[1]{#1}
\csname url@samestyle\endcsname
\providecommand{\newblock}{\relax}
\providecommand{\bibinfo}[2]{#2}
\providecommand{\BIBentrySTDinterwordspacing}{\spaceskip=0pt\relax}
\providecommand{\BIBentryALTinterwordstretchfactor}{4}
\providecommand{\BIBentryALTinterwordspacing}{\spaceskip=\fontdimen2\font plus
\BIBentryALTinterwordstretchfactor\fontdimen3\font minus \fontdimen4\font\relax}
\providecommand{\BIBforeignlanguage}[2]{{%
\expandafter\ifx\csname l@#1\endcsname\relax
\typeout{** WARNING: IEEEtran.bst: No hyphenation pattern has been}%
\typeout{** loaded for the language `#1'. Using the pattern for}%
\typeout{** the default language instead.}%
\else
\language=\csname l@#1\endcsname
\fi
#2}}
\providecommand{\BIBdecl}{\relax}
\BIBdecl

\bibitem{6G_EI}
X.~Zhu, J.~Liu, L.~Lu, T.~Zhang, T.~Qiu, C.~Wang, and Y.~Liu, ``Enabling intelligent connectivity: A survey of secure {ISAC} in 6{G} networks,'' \emph{IEEE Communications Surveys \& Tutorials}, pp. 1--1, 2024.

\bibitem{towards_6g}
Z.~Wei, F.~Liu, C.~Masouros, N.~Su, and A.~P. Petropulu, ``Toward multi-functional 6{G} wireless networks: {I}ntegrating sensing, communication, and security,'' \emph{IEEE Communications Magazine}, vol.~60, no.~4, pp. 65--71, 2022.

\bibitem{ISAC_survey}
F.~Liu, Y.~Cui, C.~Masouros, J.~Xu, T.~X. Han, Y.~C. Eldar, and S.~Buzzi, ``Integrated sensing and communications: {T}oward dual-functional wireless networks for 6{G} and beyond,'' \emph{IEEE J. Sel. Areas Commun.}, vol.~40, no.~6, pp. 1728--1767, 2022.

\bibitem{5G_6G_ISAC}
S.~Lu, F.~Liu, Y.~Li, K.~Zhang, H.~Huang, J.~Zou, X.~Li, Y.~Dong, F.~Dong, J.~Zhu, Y.~Xiong, W.~Yuan, Y.~Cui, and L.~Hanzo, ``Integrated sensing and communications: {R}ecent advances and ten open challenges,'' \emph{IEEE Internet of Things Journal}, vol.~11, no.~11, pp. 19\,094--19\,120, 2024.

\bibitem{ISAC_Caire}
Y.~Xiong, F.~Liu, Y.~Cui, W.~Yuan, T.~X. Han, and G.~Caire, ``On the fundamental tradeoff of integrated sensing and communications under {G}aussian channels,'' \emph{IEEE Trans. Inf. Theory}, vol.~69, no.~9, pp. 5723--5751, 2023.

\bibitem{V2X}
K.~Dong, M.~Mizmizi, D.~Tagliaferri, and U.~Spagnolini, ``Vehicular blockage modelling and performance analysis for mm{W}ave {V2V} communications,'' in \emph{ICC 2022 - IEEE Intern. Conf. Commun.}, 2022, pp. 3604--3609.

\bibitem{UAV_blockage}
\BIBentryALTinterwordspacing
J.~Zhao and W.~Jia, ``Channel transmission strategy for mm{W}ave hybrid {UAV} communications with blockage,'' \emph{Electronics Letters}, vol.~54, pp. 74--76, 2018. [Online]. Available: \url{https://api.semanticscholar.org/CorpusID:116845721}
\BIBentrySTDinterwordspacing

\bibitem{van_trees_BCRB}
\BIBentryALTinterwordspacing
H.~Van~Trees, \emph{Parameter Estimation {I}: {M}aximum Likelihood}.\hskip 1em plus 0.5em minus 0.4em\relax John Wiley \& Sons, Ltd, 2002, ch.~8, pp. 917--1138. [Online]. Available: \url{https://onlinelibrary.wiley.com/doi/abs/10.1002/0471221104.ch8}
\BIBentrySTDinterwordspacing

\bibitem{Poincare_dytso}
A.~Dytso, M.~Cardone, and I.~Zieder, ``Meta derivative identity for the conditional expectation,'' \emph{IEEE Trans. Inf. Theory}, vol.~69, no.~7, pp. 4284--4302, 2023.

\bibitem{WCNC_conf}
M.~B. Mohajer, L.~Barletta, D.~Tuninetti, A.~Tomasoni, D.~{Lo Iacono}, and F.~Osnato, ``{MMSE} channel estimation in fading {MIMO} {G}aussian channels with blockage: {A} novel lower bound via {P}oincar\'e inequality,'' 2024, {S}ubmitted to \emph{IEEE Wireless Communications and Networking Conference (WCNC) 2025}. Available at \url{https://arxiv.org/abs/2410.22941}.

\bibitem{Pareto_boundary}
H.~Hua, T.~X. Han, and J.~Xu, ``{MIMO} integrated sensing and communication: {CRB}-rate tradeoff,'' \emph{IEEE Trans. Wirel. Commun.}, vol.~23, no.~4, pp. 2839--2854, 2024.

\bibitem{Elements_cover}
\BIBentryALTinterwordspacing
T.~M. Cover and J.~A. Thomas, \emph{Elements of Information Theory}.\hskip 1em plus 0.5em minus 0.4em\relax John Wiley \& Sons, Ltd, 2005. [Online]. Available: \url{https://onlinelibrary.wiley.com/doi/abs/10.1002/047174882X.ch15}
\BIBentrySTDinterwordspacing

\bibitem{Capacity_bound}
T.~Marzetta and B.~Hochwald, ``Capacity of a mobile multiple-antenna communication link in {R}ayleigh flat fading,'' \emph{IEEE Trans. Inf. Theory}, vol.~45, no.~1, pp. 139--157, 1999.

\bibitem{USTM}
B.~Hochwald and T.~Marzetta, ``Unitary space-time modulation for multiple-antenna communications in {R}ayleigh flat fading,'' \emph{IEEE Trans. Inf. Theory}, vol.~46, no.~2, pp. 543--564, 2000.

\bibitem{power_constraint}
E.~Biglieri, J.~Proakis, and S.~Shamai, ``Fading channels: {I}nformation-theoretic and communications aspects,'' \emph{IEEE Transactions on Information Theory}, vol.~44, no.~6, pp. 2619--2692, 1998.

\end{thebibliography}

\end{document}